\documentclass[twocolumn,floatfix,superscriptaddress,amsmath,showpacs,showkeys,aps,prl]{revtex4}

\usepackage[final]{graphicx}
\usepackage{t1enc}
\usepackage{bm}
\begin{document}

\title{\bf Stripe-tetragonal first order phase transition in ultra-thin magnetic films}

\author{Sergio A. Cannas}
\email{cannas@famaf.unc.edu.ar}
\affiliation{Facultad de  Matem\'atica, Astronom\'{\i}a  y F\'{\i}sica, Universidad
Nacional de C\'ordoba, \\ Ciudad Universitaria, 5000 C\'ordoba, Argentina}
\altaffiliation{Member of CONICET, Argentina}
\author{Daniel A. Stariolo}
\email{stariolo@if.ufrgs.br}
\affiliation{Departamento de F\'{\i}sica,
Universidade Federal do Rio Grande do Sul\\
CP 15051, 91501--979, Porto Alegre, Brazil}
\altaffiliation{Research Associate of the Abdus Salam International
Centre for Theoretical Physics, Trieste, Italy}
\author{Francisco A. Tamarit}
\email{tamarit@famaf.unc.edu.ar}
\affiliation{Facultad de  Matem\'atica, Astronom\'{\i}a  y F\'{\i}sica,
Universidad Nacional de C\'ordoba, \\ Ciudad Universitaria,
5000 C\'ordoba, Argentina}
\altaffiliation{Member of CONICET, Argentina}

\date{\today}

\begin{abstract}
We analyze the nature of the phase transition from a smectic stripe phase to a tetragonal
phase predicted in analytic studies by Abanov et al. (Phys. Rev. B{\bf 51}, 1023 (1995)) and observed in
experiments on ultra-thin magnetic films by Vaterlaus et al.(Phys. Rev. Lett. {\bf 84}, 2247 (2000)).
At variance with existent numerical evidence, for the first time we show results of
Monte Carlo simulations
on a two-dimensional model with competing exchange
and dipolar interactions showing strong evidence that the transition is a {\em weak first
order} one, in agreement with the theoretical predictions of Abanov et al. Besides
the numerical evidence, we give further support to the first order nature of the
transition analyzing a continuum version of the model and showing that it belongs to a
large family of systems, or universality class, in which a first order transition driven
by fluctuations is expected on quite general grounds.

\end{abstract}


\maketitle

With the advances in the experimental manipulation of materials at
atomic length scales a renewed interest has grown in understanding
the thermodynamic and mechanical properties of systems such as
ultra-thin films and quasi two-dimensional magnetic materials. Part
of this interest is obviously motivated by the great amount of
potential applications they find nowadays in many different
technological fields (data storage, catalysis and electronics are
only a few examples). In this letter we are mainly concerned with
the thermodynamic properties of ultra-thin magnetic films, like
metal films on metal substrates (e.g. Fe on Cu \cite{PaKaHo1990},
Co on Au \cite{AlStBi1990}, see also Ref.\cite{DeMaWh2000} for a
recent review on the topic).

A large variety of magnetic films exhibit a spin reorientation
transition below some finite temperature $T_R$. That is, if the
magnetic film is thin enough (a few atomic layers) the atomic
magnetic moments tend to align in a direction perpendicular to the
plane of the film, because the surface anisotropy overcomes the
anisotropy of the dipolar interactions, which favors in-plane
ordering. Under these circumstances the local magnetic moments can
be regarded approximately as Ising variables. Any realistic
theoretical description of a magnetic thin film must include
long-range dipolar interactions. The competition between exchange
and dipolar interactions in these materials give rise to stable
modulated stripe-like patterns at low temperatures. In these
states the magnetic moments align along a particular axis forming
ferromagnetic stripes of constant width $h$, so that moments in
adjacent stripes are anti-aligned. Theoretical studies concluded
that the stripe state in this systems is always the most stable
one at low enough temperatures, provided that the exchange
parameter exceeds some small positive critical
value\cite{AbKaPoSa1995,MaWhRoDe1995}. Calculations based on a
continuum approximation by Abanov et al. \cite{AbKaPoSa1995}
predicted that, before reaching the paramagnetic state at high
temperatures, the stripe phase undergoes a transition into a phase
characterized by domains with predominantly square corners, that
they called a {\it tetragonal liquid}. They also concluded that
the stripe-tetragonal liquid  transition should be either first
order or the two phases might be separated by a third phase
characterized by rotational domain walls defects, that they called
an {\it Ising nematic phase}. Monte Carlo calculations on the
square lattice carried out by Booth et al.~\cite{BoMAWhDe1995}
confirmed the presence of the tetragonal liquid phase, but they
did not find any evidence of the Ising nematic phase. However, the
results of Booth et al appeared to be consistent with a {\it
continuous} transition rather that a first order one, as could be
expected from Abanov et al theoretical results. Nevertheless,
Booth et al pointed out that the possibility of a weak first order
transition cannot be excluded on the base of their Monte Carlo
simulations. Recent imaging studies using scanning electron
microscopy with polarization analysis on ultra-thin films of fcc Fe
on Cu(100) verified the existence of the tetragonal liquid
phase\cite{VaStMaPiPoPe2000,PoVaPe2003}.
 No evidence was found of an intermediate
nematic phase of the type predicted by Abanov et al.\cite{AbKaPoSa1995}
 in the transition
from the stripe phase to the tetragonal liquid. The thermodynamic nature
of the stripe-tetragonal transition could not be determined by the imaging technique and
this question remains unanswered.
In this letter we present both   Monte Carlo (MC) and analytical calculations that provide
new evidence that the stripe-tetragonal liquid transition is indeed a weak
fluctuation-induced first order one.


We consider a system of magnetic dipoles on a square lattice in which the magnetic moments
 are oriented perpendicular to the plane of the lattice, with both nearest-neighbor
ferromagnetic exchange interactions and long-range dipole-dipole interactions between
moments. The thermodynamics of this system is ruled by the dimensionless
Hamiltonian~\cite{difference}:
\begin{equation}
{\cal H}= - \delta \sum_{<i,j>} \sigma_i \sigma_j + \sum_{(i,j)}
\frac{\sigma_i \sigma_j}{r^3_{ij}}
\label{Hamilton1}
\end{equation}
\noindent where $\delta$ stands for the ratio between the exchange
$J_0>0$ and the dipolar $J_d>0$ interactions parameters, i.e.,
$\delta = J_0/J_d$. The first sum runs over all pairs of nearest
neighbor spins and the second one over all distinct pairs of spins
of the lattice; $r_{ij}$ is the distance, measured in crystal
units, between sites $i$ and $j$. The energy is measured in units
of $J_d$. The overall (known) features of the equilibrium phase
diagram of this system can be found in
Refs.\cite{DeMaWh2000,MaWhRoDe1995,BoMAWhDe1995,GlTaCa2002}, while
several dynamical properties at low temperatures can be found in
Refs.\cite{SaAlMe1996,ToTaCa1998,StCa1999,GlTaCaMo2003}. The
threshold for the appearance of the stripe phase in this model is
$\delta_c=0.425$~\cite{MaWhRoDe1995,difference}.

We performed extensive MC simulations of the  model defined by
Eq.(\ref{Hamilton1})   on a square lattice with $N=L^2$  sites,
for  systems sizes ranging from $L=16$ to $32$ using the
Metropolis algorithm. Periodic boundary conditions were
implemented using  the Ewald summation technique. Although we
obtained results for different values of $\delta$ ranging from $1$
to $3$, most of the numerical work was focused on $\delta=2$
(corresponding to a low temperature stripe phase  of width $h=2$),
for which the first order nature of the transition is more clearly
defined. To estimate both equilibration and decorrelation times we
analyzed the behavior of the  two-times correlation function  for
different system sizes and temperatures.
\begin{figure}
\begin{center}
\includegraphics[width=8cm,height=4.5cm,angle=0]{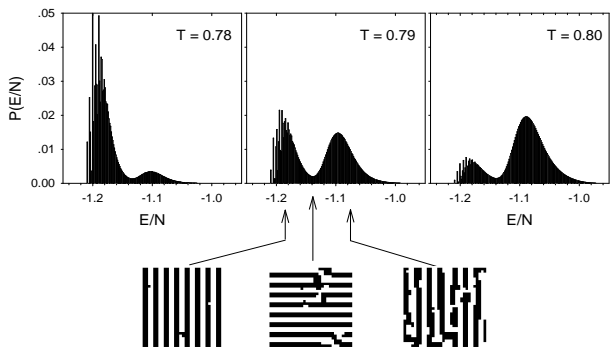}
\caption{\label{fig1}Energy per spin histograms for $\delta=2$,
$L=32$ and different temperatures around the critical one $T_c
\approx 0.79$. The images below the histogram illustrate some
typical equilibrium spin configurations corresponding to the
maxima and the minimum at the critical temperature.}
\end{center}
\end{figure}
 After an equilibration period
of up to $10^5$ Monte Carlo Steps (MCS), every data point was
calculated over a large single run for periods  ranging from
$2\times 10^6$ MCS for the smallest size, up to $4\times 10^7$ MCS
for the largest one ($L=32$). To locate the transition and
characterize  its order we calculated the energy per spin
histograms for different temperatures, from which we obtained both
the specific heat $C_L = \frac{1}{NT^2} \left( \left< H^2 \right>
-\left< H \right>^2\right)$ \noindent and the fourth order Binder
cumulant\cite{ChLaBi1986} $V_L = 1- \left< H^4 \right>/3\left< H^2
\right>^2$.
 The typical behavior of the energy histograms  is shown
in Fig.\ref{fig1} for $\delta=2$ and $L=32$. The double peak
structure is characteristic of a first order phase
transition\cite{ChLaBi1986,LeKo1991}.
  We observed that the measurement time (in MCS) needed to sample
the energy distribution effectively increases exponentially with
the linear system size $L$. This is because the system spends
large periods of time in each one of the two phases, with a mean
life time in every phase that increases with $L$, consistently
with the expected\cite{LeKo1991} increase in the energy free
barrier between phases $\Delta F(L) \sim L^{d-1}$. A careful
inspection of the typical equilibrium spin configurations
associated with every phase (i.e.,  those with energies around the
maxima of the energy distribution) shows indeed that the high
temperature phase presents a tetragonal structure similar to that
found by Booth et al. \cite{BoMAWhDe1995} (see Fig.\ref{fig1}),
while the low temperature phase is the $h=2$ stripe one (also
 notice the coexistence of domains of both phases at energies
corresponding to the minimum of the histogram). Moreover, a MC
calculation of the structure factor $S({\bf k}) =
\left<\left|\sum_i \sigma_i e^{i {\bf k.r_i}} \right|^2 \right>$
at temperatures above, but near the critical one, show a very
similar shape to that observed by MacIsaac et al.
\cite{MaWhRoDe1995} at higher values of $\delta$: four sharp peaks
symmetrically placed on the two principal axes of the Brillouin
zone, which characterize the tetragonal structure
\cite{DeMaWh2000}. As the temperature increases the peaks become
smeared into a circle with the shape of a four-peaked crown that
gradually disappears. This indicates a continuous loss of the
fourfold symmetry as the systems becomes paramagnetic. For large
values of $\delta$ ($\delta\geq 3$) this transition is also
reflected in the presence of a secondary peak at high temperatures
in the specific heat, which does not depend on the system size
\cite{BoMAWhDe1995}, indicating that the transition
 does not have an associated singularity in
the thermodynamic potentials. As already observed by Booth et al
\cite{BoMAWhDe1995}, this secondary peak becomes less pronounced
as $\delta$ decreases; we observed that for values of $\delta\leq
2.5$ it becomes indistinguishable. However, the presence of the
transition remains clearly detectable in the behavior of the
structure factor.

\begin{figure}
\begin{center}
\includegraphics[width=8cm,height=5.5cm,angle=0]{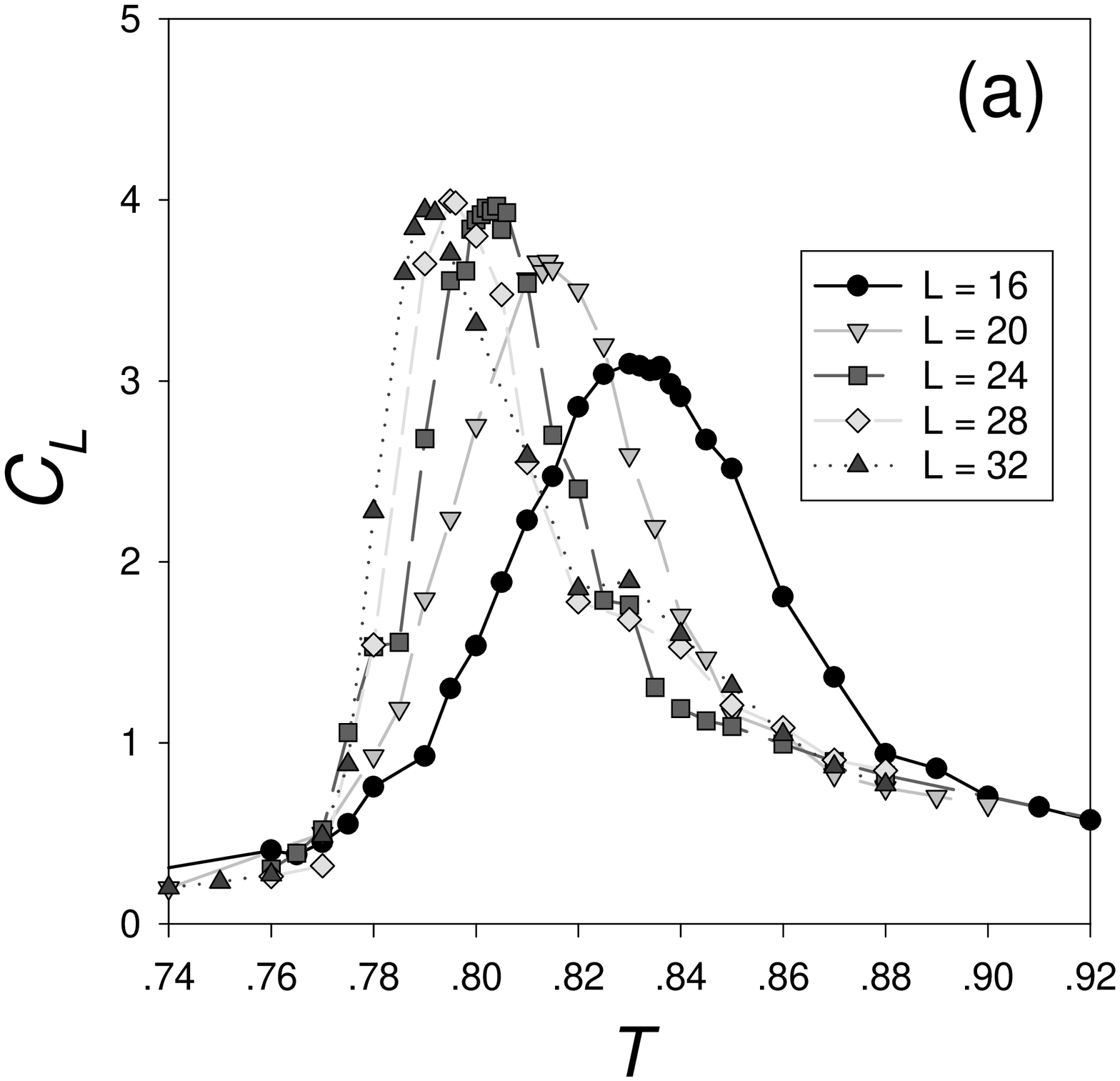}
\includegraphics[width=7cm,height=5.5cm,angle=0]{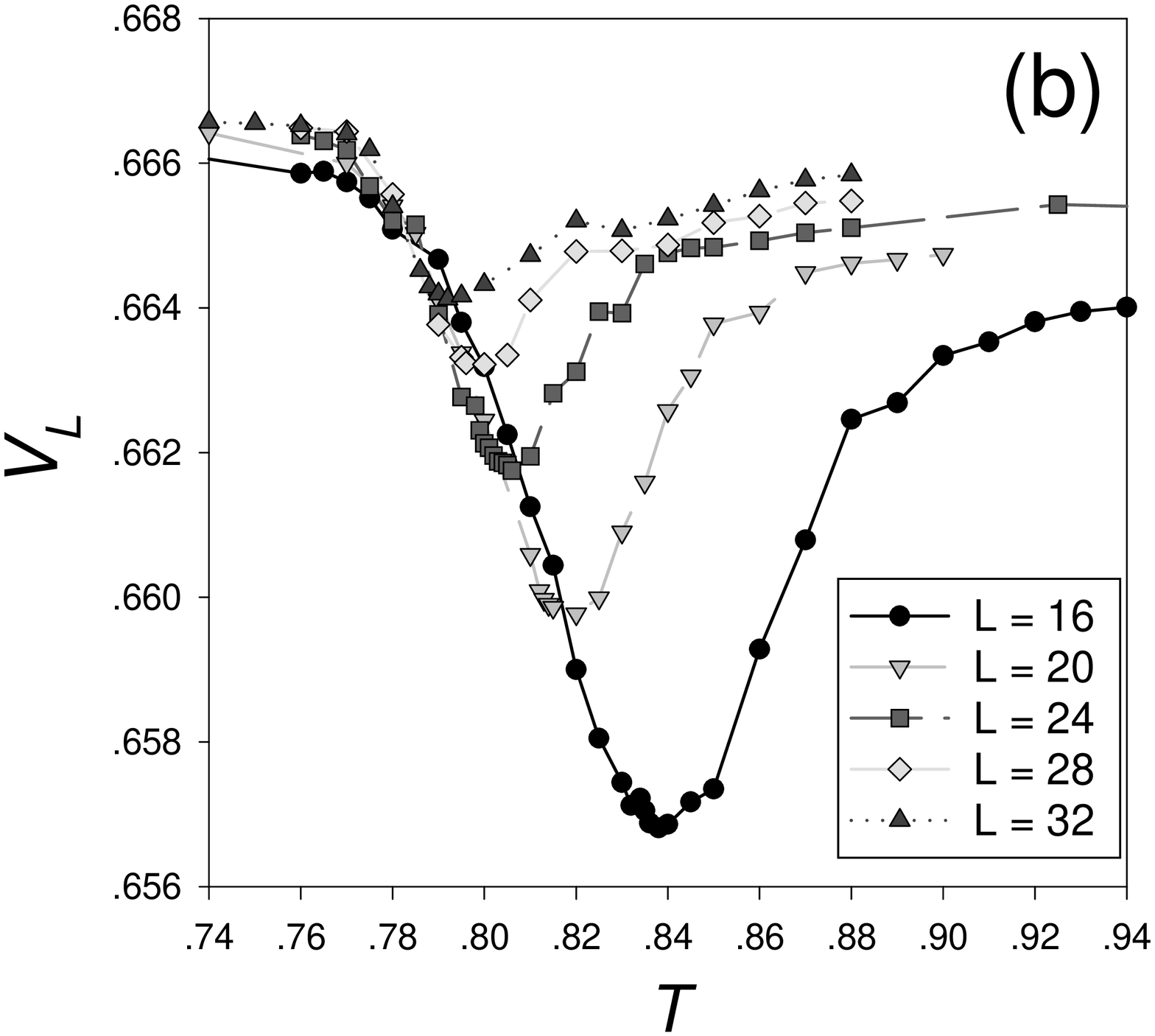}
\caption{\label{fig2}MC calculations for $\delta=2$ and different system sizes $L$. (a)
Specific heat $C_L$ vs. $T$; (b) Binder cumulant $V_L$ vs. T.}
\end{center}
\end{figure}

\begin{figure}
\begin{center}
\includegraphics[width=7cm,height=5cm,angle=0]{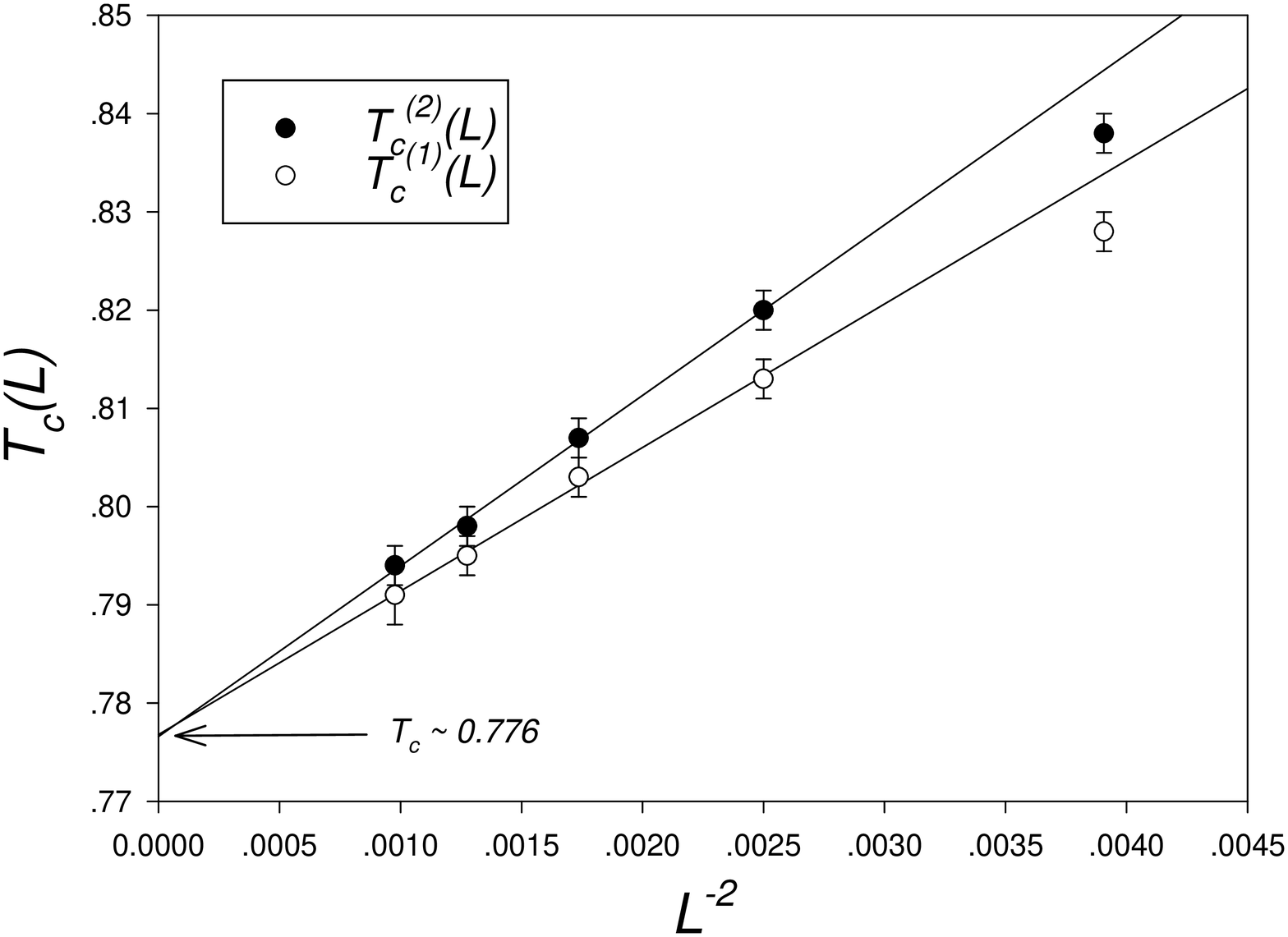}
\caption{\label{fig3}Pseudo critical temperatures $T_c^{(1)}$ (maximum of the specific
heat) and $T_c^{(2)}$ (minimum of the Binder cumulant) vs. $L^{-2}$.}
\end{center}
\end{figure}

\begin{figure}
\begin{center}
\includegraphics[width=8cm,height=3.3cm,angle=0]{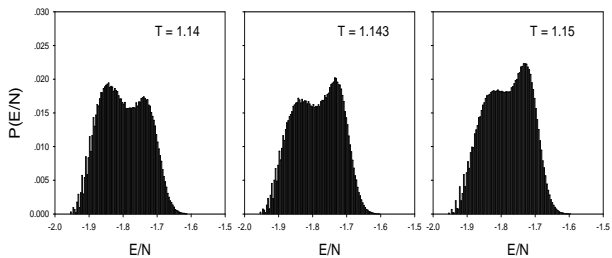}
\caption{\label{fig4}Energy per spin histograms for $\delta=2.6$,
$L=32$ and different temperatures around the critical one.}
\end{center}
\end{figure}

The temperature variation of the specific heat and the Binder
cumulant for $\delta=2$ and various system sizes is shown in
Fig.\ref{fig2}.
 We see that the location of the maximum of the
specific heat and the minimum of the Binder cumulant shift in a
size dependent fashion at pseudo-critical temperatures
$T_c^{(1)}(L)$ and $T_c^{(2)}(L)$ respectively. Both quantities
are plotted vs $L^{-2}$ in Fig.\ref{fig3}, showing the expected
finite size scaling behavior for a temperature-driven first order
phase transition\cite{ChLaBi1986,LeKo1991} $T_c^{(1)}(L)\sim T_c
+AL^{-d}$ and $T_c^{(2)}(L)\sim T_c +BL^{-d}$ with $B>A$, where
$T_c$ is the transition temperature of the infinite system. Note
that the internal energies of both phases (corresponding
approximately to the energies of the maxima of the energy
distribution) are located very close to each other. This property
is also reflected in the rather shallow shape of the minimum of
the Binder cumulant, evidencing the weak nature of the transition.
We observed that these effects become more pronounced as $\delta$
increases, with the internal energies of both phases approaching
continuously to each other. For values of $\delta>2.6$ the double
peak structure of the energy histogram (together with the minimum
of the Binder cumulant) seems to disappear, or at least it becomes
undetectable for small system sizes, as can be appreciated in
Fig.\ref{fig4} for $\delta=2.6$. We see that the internal energies
of both phases (roughly corresponding to the energies of the
maxima of the histogram) near the transition approach to each
other as $\delta$ increases.  This fact explains the seemingly
continuous nature of the transition observed by Booth et al., whose
calculations were performed for $\delta\geq 3$
\cite{BoMAWhDe1995}. Clearly, considerably larger systems must be
simulated in order to get reliable data for larger $\delta$. A
similar effect is observed as $\delta$ decreases approaching
$\delta_c$, where the double peak structure of the histogram also
seems to disappear at all. Hence, the numerical data show an
optimal value of $\delta$ around $\delta=2$, where the first order
transition becomes strongest, that its, with the largest latent
heat. Since strong finite size effects are always expected in
systems with long range interactions, numerical simulations in
considerably larger systems are also required in this case to get
a more clear picture about this phenomenon. However, it is worth
mentioning that some numerical simulations for larger system sizes
($L=48$ and $L=64$) showed that the double peak structure of the
energy histogram persists and becomes more pronounced as $L$
increases. We now turn our attention to an analytic approximation
which gives us some insight in the expected outcome of such
simulations for the other regions of the phase diagram.

It was recognized long ago~\cite{Br1975,HoSw1995} that systems in which the spectrum
of  fluctuations has a minimum at a non-zero wave vector can undergo a first order
transition driven by fluctuations, in contrast to the second order transition predicted by
mean field for this
 kind of systems. Since the original work by Brazovskii, the proposed scenario was
shown to describe correctly the phase transitions present in a
large variety of systems like cholesteric liquid crystals, the
nematic to smectic-C transition, pion condensates in neutron
stars, onset of Rayleigh-B\'enard convection and micro-phase
separation in diblock copolymers ~\cite{HoSw1995, SeAn1995}. More
recently the Brazovskii scenario has been successfully applied to
the analysis of the phase
 transition between
the disordered and modulated phases in three dimensional systems with attractive short
range
interactions and repulsive long range  Coulomb interactions~\cite{GrKrTaVi2002,ViTa1998}.
In spite of its success and ubiquity in correctly describing the physics behind a phase
transition
in systems with competing interactions, the Brazovskii
scenario was almost not considered in relation with ultra-thin films and dipolar systems
~\cite{GaDo1982}.
Indeed we have verified that a continuum version of Hamiltonian (\ref{Hamilton1})
presents
a fluctuation
induced first order phase transition for any value of $\delta$. The starting point is a
Landau-Ginzburg
functional which in Fourier space has the form:
\begin{widetext}
\begin{equation}
{\cal H}= \frac{V}{2} \int \frac{d{\vec k}}{(2\pi)^2} A(k) \phi({\vec k}) \phi(-{\vec
k})
+ \frac{uV}{4} \int \frac{d{\vec k_1}}{(2\pi)^2} \frac{d{\vec k_2}}{(2\pi)^2}
\frac{d{\vec k_3}}{(2\pi)^2}
\phi({\vec k_1}) \phi({\vec k_2}) \phi({\vec k_3}) \phi({-{\vec k_1}-{\vec k_2}-{\vec
k_3}})
\end{equation}
\end{widetext}
\noindent where the spectrum of fluctuations is given by $A(k)= r_0+k^2+J(k)/\delta $
and $r_0$ is proportional to the reduced temperature near the critical point of the
mean field approximation.
In our case the Fourier transform of the dipolar interaction is
$J(k) = \mbox{ }_1F_2\left( \{-1/2\},\{1/2,1\},-k^2/4\right)-k$,
 where $\mbox{ }_1F_2\left( \{\alpha\},\{\beta_1,\beta_2\},z\right)$ is a
generalized hyper-geometric series~\cite{GrRi1994}. The spectrum of
fluctuations presents a minimum at a nonzero $k_{min}=k_0$. This
means that, for $d \geq 2$  the spectrum is minimized in a
spherical shell in Fourier space. When fluctuations of the order
parameter are considered self-consistently, the degeneracy of the
minimizing vector will make the disordered phase meta-stable for
any finite temperature and drive the transition to first order.
More explicitly, within a Hartree approximation the correlation
function of the fluctuations in the disordered phase is given by:
\begin{equation}
C^{-1}(\vec{k}) = A(k) + 3u \int \frac{d{\vec q}}{(2\pi)^2} C(\vec{q})
\end{equation}
\noindent Then, the renormalized mass $r = r_0 + 3u \int \frac{d{\vec q}}{(2\pi)^2}
C(\vec{q})$ is given by:
\begin{equation}
r = r_0 + 3u \int \frac{d{\vec q}}{(2\pi)^2} \frac{1}{r+q^2+J(q)/\delta}
\end{equation}
\noindent The point of absolute instability of the disordered solution occurs at $r =
-k_0^2-J(k_0)/\delta $.
It is easy to check that this point is only reached for $r_0 \to -\infty$ (i.e. for $T\to
0$), implying that
 the disordered phase never loses stability.  At a finite
temperature a modulated phase with a finite amplitude given by
$m^2=(r_m+k_0^2+J(k_0)/\delta )/3u$
 appears, where the renormalized mass in the modulated phase $r_m$ is given
self-consistently
by:
\begin{eqnarray}
-r_m &=& r_0 + 3u \int \frac{d{\vec q}}{(2\pi)^2} \frac{1}{r_m+q^2+J(q)/\delta}\nonumber
\\
  &&       + 2\left( k_0^2 + \frac{J(k_0)}{\delta} \right)
\end{eqnarray}
In a region $r_0 \leq r_0^*(\delta)$ a real solution to the above equation exists. If a
point
$r_0^c < r_0^*$ exists where the free energies of the modulated and disordered solutions
cross
each other, then a first order transition driven by fluctuations appears.
Following Brazovskii~\cite{Br1975} we have determined the point $r_0^c(\delta)$ where
the free energy difference
\begin{widetext}
\begin{equation}
u\Delta F = \int_r^{r_m} dr'  \left( \frac{1}{6}+\frac{u}{2}
\int \frac{d{\vec q}}{(2\pi)^2} \frac{1}{(r'+q^2+J(q)/\delta)^2}\right) \left(
\frac{r'+r_0}{2}
+ k_0^2 + \frac{J(k_0)}{\delta} + \frac{3u}{2}
\int \frac{d{\vec q}}{(2\pi)^2} \frac{1}{r'+q^2+J(q)/\delta} \right)
\end{equation}
\end{widetext}
\noindent changes sign. It is possible to show that for {\em any} $\delta > 0$ a first
order transition appears. The transition line $r_0^c(\delta)$ is a monotonously increasing
function of $\delta$.
%

In summary, we have shown strong numerical evidence that the phase
transition between
the low temperature stripe phase and the tetragonal phase observed in ultra-thin
magnetic films is a weak first order transition. Moreover, to the extent to which a
continuum Hartree approximation of the model simulated is valid, the transition is
a first order one driven by fluctuations, at variance with the second order nature
predicted by mean field theory. It would be interesting to perform simulations in larger
lattices and with smarter Monte Carlo algorithms in order to analyze if the first order
nature extends to a wider region of the phase diagram as implied by the continuum model.
The challenge persists to probe the nature of this transition experimentally
~\cite{VaStMaPiPoPe2000,PoVaPe2003}.

\begin{acknowledgments}
This work was partially supported by grants from Consejo Nacional
de Investigaciones Cient\'\i ficas y T\'ecnicas CONICET
(Argentina), Agencia C\'ordoba Ciencia (C\'ordoba, Argentina),
Secretar\'{\i}a de Ciencia y Tecnolog\'{\i}a de la Universidad
Nacional de C\'ordoba (Argentina), Funda\c{c}\~ao VITAE (Brazil)
and CNPq (Brazil). We wish to thank R. Díaz-Méndez and Roberto
Mulet for their help in the implementation of the Ewald sums.
\end{acknowledgments}

\end{document}